\documentclass[12pt]{article}

\usepackage{amssymb,amsmath, epsfig}
\usepackage{changepage}
\usepackage{caption}
\usepackage[section]{placeins}
\usepackage{rotating}
\makeatletter
\AtBeginDocument{%
  \expandafter\renewcommand\expandafter\subsection\expandafter{%
    \expandafter\@fb@secFB\subsection
  }%
}
\makeatother

\def\p{\partial}

\def\nh{\noindent\hangindent=1 true cm \hangafter = 1}

\def\nh{\noindent\hangindent=1 true cm \hangafter = 1}

\def\B {\begin{eqnarray*}}

\newcommand{\bel}[1]{\begin{equation}\label{#1}}

\newcommand{\be}{\begin{equation}}
\newcommand{\qe}{\end{equation}}
\newcommand{\ee}{\end{equation}}
\newcommand{\baS}{\begin{eqnarray}}
\newcommand{\ba}{\begin{eqnarray}}

\newcommand{\ea}{\end{eqnarray}}

\def\EN{\end{eqnarray*}}

\begin{document}
\title{Gompertzian  population growth under some deterministic and stochastic jump schedules\\
\   \\
{\normalsize
Henry C. Tuckwell$^{1,2\dagger, *}$\\   \
\  \\ 
$^1$ School of Electrical and Electronic Engineering, University of Adelaide,\\
Adelaide, South Australia 5005, Australia \\
 \              \\
$^2$ School of Mathematical Sciences, Monash University, Clayton,
Victoria 3800, Australia \\
\  \\
$^{\dagger}$ {\it Email:} henry.tuckwell@adelaide.edu.au\\ 
\     \\}}

\maketitle

%
%
\begin{abstract}  
Many cell populations, exemplified by certain tumors, grow 
approximately according to a Gompertzian growth model which
has a slower approach to an upper limit than that of logistic growth.
Certain populations of animals and other organisms have also
recently been analyzed with the Gompertz model. 
This article addresses the question of how long it takes to 
reduce the population from one level to a lower one under a schedule  of  sudden decrements, each of which removes a given fraction of the cell mass or population. A deterministic periodic schedule is first examined and 
yields exact results for the eradication or extinction time which is defined as that
required to reduce the number of cells to less than unity. 
The decrements in cell mass at each hit could correspond to an approximation to reduction of a tumor by external beam radiation therapy. 
The effects of variations in magnitude of successive decrements,
the time interval between them, the initial population size and the intrinsic growth rate are calculated and results presented graphically. 

 With a schedule governed by a Poisson process, the number of organisms or cells satisfies a stochastic differential equation whose solution sample paths have downward jumps
as random times. The moments of the exit time then satisfy a system of recurrent differential-difference equations.   A simple transformation results in a simpler system which has been studied both analytically and numerically in the context of interspike intervals of a model neuron. Results are presented for the mean eradication time for various frequencies and magnitudes of hits and for various eventual and initial population sizes. The standard deviation of the eradication time is also investigated.

 \end{abstract}
\centerline {{\it Keywords}: Gompertz model; tumor; eradication time;
jump process; external beam radiation therapy}

\newpage
\rule{60mm}{1.5pt}
\tableofcontents
\rule{60mm}{1.5pt}
%
  \section{Introduction}
An approximate empirically based model for the growth of some 
biological populations
is provided by the Gompertzian function first
introduced as a model for human mortality (Gompertz, 1825).
An early review of its application to growth curves using  differential equations was that of Winsor (1932).

Starting in the 1960s and 1970s, the Gompertz model 
 became popular for its ability to describe the growth of certain tumor cell populations (Laird, 1964; Simpson-Herren and Lloyd, 1970; Smith and Tuckwell, 1974; Norton et al., 1976). This application has continued to 
the present day. Norton (1988), for example,  applied the model to breast cancer.  
Although the Gompertzian description of tumor growth  is purely phenomenological, it can provide a useful simplification which avoids taking into account  geometric factors, vascularization, cell types and details of the cell cycles, which require a large number of parameters (Burton, 1966; Jansson and
Revesz, 1974; Dibrov et al., 1985;  Bajzer et al., 1997).
Recently there has appeared an interesting comparison of the predictions of seven ordinary differential equation models, including the Gompertzian, for the growth of tumors (Murphy et al., 2016). It was found that wide discrepancies arise in the application of these models which
has implications for the choice of  suitable doses of chemotherapeutical agents. Behera and O'Rourke (2008) analyzed the Gompertzian model
for tumor growth with additive and multiplicative noise.  
The familiar logistic model, included in the seven models, 
sometimes outperforms other models (Vaidya et al., 1982) and has also
been analyzed with additive and multiplicative white noises (Ai et al., 2003). 

The important issue of modeling the responses of tumors to radiation therapy and chemotherapy has been addressed by several authors. The targets of both forms of treatment is the destruction of the DNA of tumor cells, rendering them incapable of mitosis.  Depending on the nature of the treatment a different mathematical formulation is employed. Chemotherapy, applied systemically, usually involves a 
continuous negative effect on growth (Sachs et al.,2001). The most common form of radiation therapy with an  external beam is performed at regular time intervals, often daily for 5 days per week, hence resulting in
 discrete and sudden declines in tumor size  (Rockne et al., 2009). 
However tumor size is not a very useful variable for many tumors  because dead cells
tend to remain in place, at least in the short term.  For small tumors in their initial stages,  exponential models may be employed (Badri et al., 2016) rather than
Gompertzian or other saturating processes. Rockne et al. (2009)
explored a spatio-temporal model of tumor growth in the form
of a reaction-diffusion system. 

The Gompertz model has also often been employed for populations of diverse organisms, one of the first applications being to the human population of the United States (Davies, 1927).  Other examples include 
plant disease (Berger, 1981), several mammalian populations  (Zullinger et al., 1984),
sage-grouse in North America (Beever and Aldridge, 2011), fish and insects (Eberhardt and Breiwick, 2012) and ungulates (Ferguson and Ponciano, 2015). 

\section{Description of model}

 In this article we only study the effects of density independent 
  ``disasters'' on Gompertzian growth. The deterministic differential
equation describing such growth in the unimpeded case for a population of size $N(t)$ at time $t$ contains only two parameters:
\be \frac{dN}{dt}=rN(\ln K - \ln N), N(0)=N_0 \in (0,K), \ee
where $r$ is the intrinsic growth rate and $K$ is the asymptotic 
maximum population size (carrying capacity) when $r$ is positive. 
Equation (1) can also be written as a system of two differential equations (Simpson-Herren and Lloyd, 1970).  The solution of (1) is 
\be N(t)=K \exp\bigg\{\ln\bigg[ \frac{N_0}{K} \bigg] \exp(-rt)\bigg\}, \ee
which shows the slow approach to $K$ as $t \rightarrow \infty$.

In the context of populations of organisms, it may be assumed that
 sudden decrements can occur due to accidents,
external attack by hostile species, disease outbreaks, floods or fires etc.
Such decrements are usually unpredictable. 
 When tumors are subjected to certain clinical treatments, such as regimes of external radiation therapy, then
each successful treatment will result in a decline in the number of viable tumor cells. 

 In what follows we assume that decrements are proportional to the number of organisms or cells present and that they are rather sudden so that they may be described as downward jumps or discontinuities.  Such
a dynamic has been depicted in certain chemotherapeutic regimes for tumors by Aroesty et al. (1973) (c.f. Sachs et al., 2001). 
In  such cases a  treatment at time $t$ gives, in the manner of a disaster
in a population of organisms, 
\be N(t^+) = N(t^-) - \epsilon N(t^-), \ee
where $\epsilon$ is a positive constant.   A more realistic representation
would have $\epsilon$ as a random variable but this complication is
ignored here. Another possibility, more appropriate for a widely dispersed population,  is that the decline is not
proportional to the number of cells present but rather is a random
number of cells also described by a (constant) random variable,
\be   N(t^+) = N(t^-) - \epsilon. \ee
However,  analysis is able to be performed more readily if the assumption of (3) is made and this case seems to be of practical interest. 

\section{Deterministic regime}
If one assumes that sudden decrements proportional to existing
population size occur periodically then the time course of the
population can be found analytically, as was observed by Aroesty et al. (1973) in connection with chemotherapy for tumors.  Let  the population evolve according to 
\be \frac{dN}{dt}=rN(\ln K - \ln N)-N\sum_{i=1}^{\infty} \epsilon \delta(t-iT), N(0)=N_0 \in (0,K), \ee
so that the population jumps downward by a fraction $\epsilon$ of its
current size at intervals of $T$. Putting
\be k=\ln K, Y=k-\ln N,  \alpha = \ln \frac{1}{1-\epsilon} \ee
yields the linear differential equation
\be \frac{dY}{dt}=-rY + \alpha \sum_{i=1}^{\infty} \delta(t-iT).\ee
If the initial value of $Y$ is $Y_0$, then after $m$ downward jumps in $N$
have occurred, 
\be Y(mT^+)=Y_0e^{-mrT} + \alpha S_m, \ee
where $S_m$ is a geometric series
\be S_m=1 + e^{-rT} + \dots +e^{-(m-2)rT}       + e^{-(m-1)rT}   \ee
which sums to give
\be Y(mT^+)=Y_0e^{-mrT} + \alpha \frac { e^{-mrT}-1}{e^{-rT}-1}. \ee
The pre-jump value is
\be Y(mT^-)=Y(mT^+) - \alpha. \ee
The asymptotic large time value post jump value is
\be Y^+(\infty) = \frac {\alpha} {1 - e^{-rT}} \ee
with corresponding value for $N$ of
\be N^+(\infty) = \exp\bigg(k -  \frac {\alpha} {1 - e^{-rT}} \bigg).  \ee
If extinction is defined as achieving a level less than 1, then
in terms of the original parameters the condition for extinction 
is
$$ \frac { \ln (1-\epsilon)} {e^{-rT}-1}  > \ln K. $$
If this condition is met, 
the actual time to extinction $T_{ext}$ is more easily expressed in terms of $Y$. The time taken for $N$ to first become less than 1 is the same as that for $Y$ to exceed $k$ so 
$$ T_{ext}= T \inf \{m|Y_0e^{-mrT} + \alpha \frac { e^{-mrT}-1}{e^{-rT}-1} >k \}. $$
However when exploring extinction it is more simply done numerically.
  \subsection{Numerical results}
A few examples of trajectory values pre- and post-jump were calculated
for $Y$ using the above formulas and converted to values for $N$ using
\be N(t)=e^{k-Y}. \ee 
The results are displayed in Figures 1 to 4, where only values of $N$ at the pre and post jump values are shown, not the trajectories between jumps.  The following set of parameter values  was taken as a standard set $K=e^{20}= 4.8517 \times 10 ^8$, $T=2$, $\alpha=0.5$,  $N_0=0.9K$, $r=0.02$, 
together with 15 jumps so that the number of points is 31. The value 
$\alpha=0.5$ is equivalent to $\epsilon=1-\exp(-\alpha)= 0.3935$ in terms of jumps in $N$. 

Results for the standard set are included as blue curves in all of the Figures 1 to 4. In each of the 4 Figures one parameter is varied, 
being $\alpha$, $T$, $N_0$ and $r$ respectively.  In Figure 1, the 
decline in $N$ is sharp, being roughly exponential, for the standard set,
but when $\alpha$ is reduced the decline is much slower and to a larger value as predicted by Equ. (13).

 Figure 2 shows very little difference when $T$ is doubled from 2 to 4 but a substantial difference in both the rate of decline and the asymptotic value when $T=10$. Figure 3 displays  
results for various $N_0$, but truncated to emphasize the large-time
declines. The latter reveal little dependence on $N_0$ as would be expected from Equ. (13).  Changes in $r$ from the standard value
give results as expected with a much higher asymptotic value attained much sooner for the largest value $r=0.2$. Results such as these may be of some indicative utility in predicting the effects of various regimes of radio or chemotherapy 
on tumor reduction. 

     \begin{figure}[!h]
\begin{center}
\centerline\leavevmode\epsfig{file=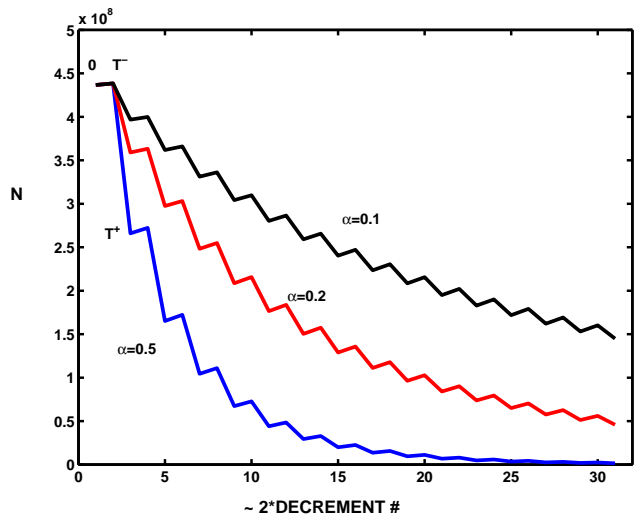,width=3.25in}
\end{center}
\caption{Time course of Gompertzian growth with periodic decrements
and various $\alpha$ as indicated.
Values of $N$ are shown just prior to (e.g., $T^-$) and just after
(e.g., $T^+$) each decrement. } 
\label{fig:wedge}
\end{figure}

     \begin{figure}[!h]
\begin{center}
\centerline\leavevmode\epsfig{file=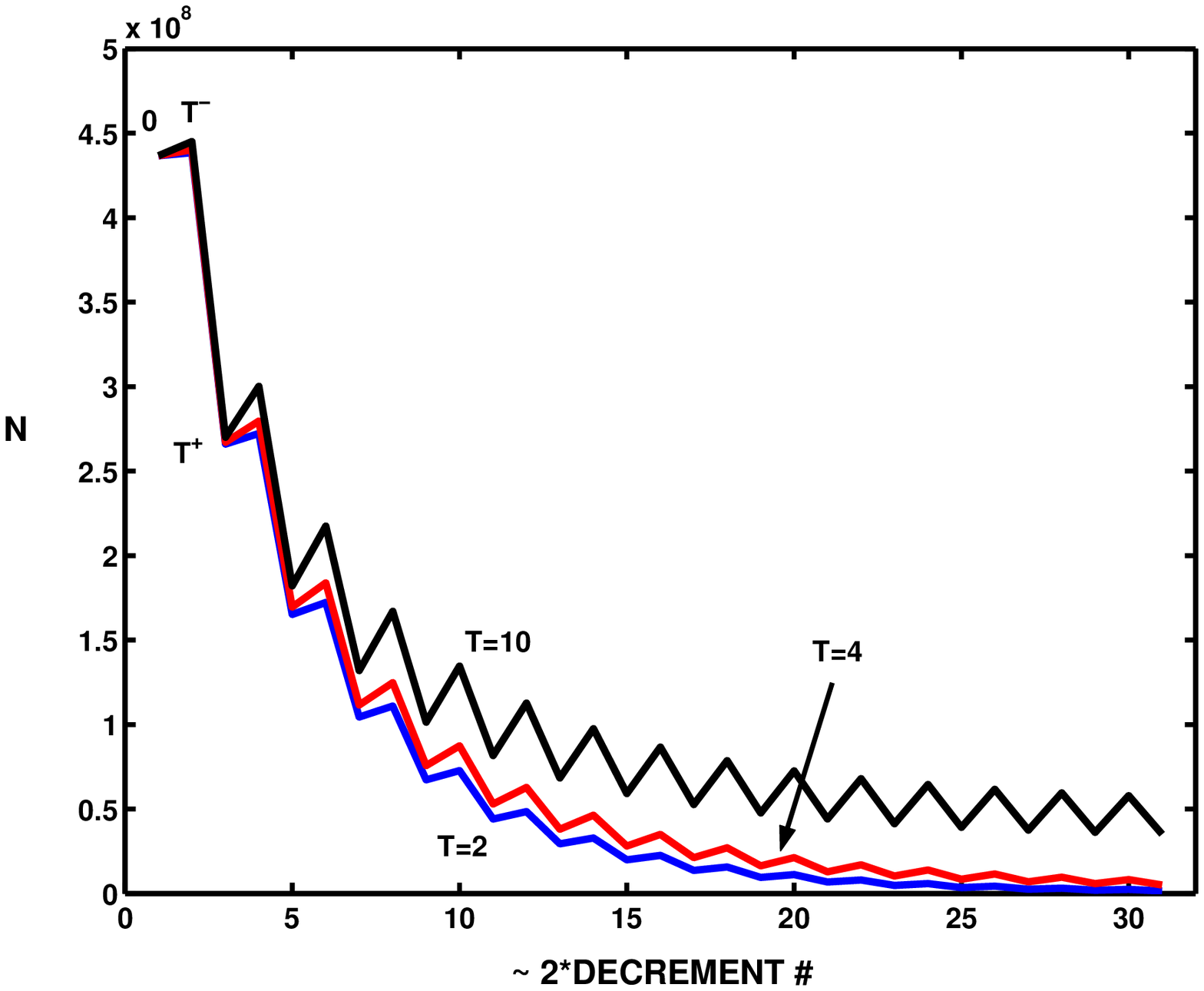,width=3.25in}
\end{center}
\caption{Time course of Gompertzian growth with periodic decrements
and various $T$ as indicated.
Values of $N$ are shown just prior to and just after each decrement. } 
\label{fig:wedge}
\end{figure}

     \begin{figure}[!h]
\begin{center}
\centerline\leavevmode\epsfig{file=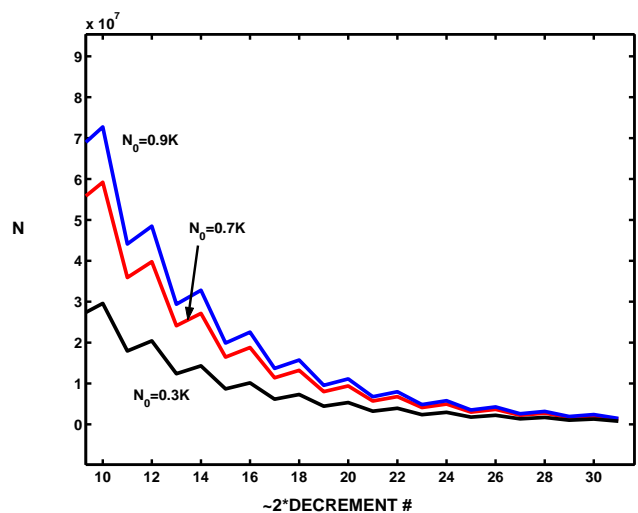,width=3.25in}
\end{center}
\caption{Time course of Gompertzian growth with periodic decrements
and various initial sizes $N_0$ as indicated.
Values of $N$ are shown just prior to and just after each decrement.} 
\label{fig:wedge}
\end{figure}

\section{Stochastic regime}
There are many possible ways to incorporate randomness into
models of population growth. One useful and popular method is
through stochastic differential equations (SDEs) for Markov processes.
Often such processes are continuous and represented by diffusion processes (Tuckwell, 1974, Tuckwell and Le Corfec, 1998, for examples), for which a paradigm in the temporarily homogeneous case is
\be dN=f(N)dt + g(N)dW \ee
where $f$ and $g$ are suitable functions (see for example Gihman and
Skorohod, 1972; Oksendal, 2000) and $W$ is a 1-parameter standard
Wiener process or Brownian motion with mean zero and $Var(W(t))=t$.

     \begin{figure}[!h]
\begin{center}
\centerline\leavevmode\epsfig{file=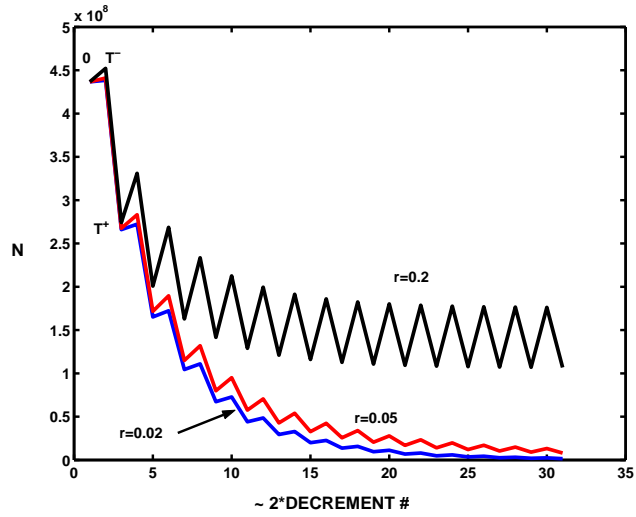,width=3.25in}
\end{center}
\caption{Time course of Gompertzian growth with periodic decrements
and various growth rates $r$ as indicated.
Values of $N$ are shown just prior to and just after  each decrement.} 
\label{fig:wedge}
\end{figure}

\subsection{Theory}
Solutions of equations like (15) do not contain the possibility of sudden 
large (discontinuous) changes in $N$ which may occur in real 
populations. However, such large random fluctuations can be 
incorporated by considering more general stochastic differential equations of the form
\be dN=f(N)dt + g(N)dW + \int_R h(N,u)n(dt,du), \ee
where $n(t,A)$ for $A \in B(R) $ is a Poisson process (the number of
jumps with magnitudes in the Borel set A up to time t) such that if
$E[n(t,A)]=tM(A)$ then
\be Pr[n(t,A)=k]= \frac {(M(A)t)^k \exp(-M(A)t)}{k!}, \ee
for $k=0,1,2,... $. 
The theory of such processes was developed by Feller (1940) for a process without a diffusion component and in the more general case
by Ito (1951), Skorohod (1965) and Gihman and Skorohod (1972). 
In the last of these references it is shown how the forward and backward
Kolmogorov equations for the transition probability density function can
be written down for a process satisfying an SDE of the form of (16). In general such equations are functional partial differential equations and are more difficult to solve than the corresponding equations for the simpler case in which a diffusion process is unaccompanied by a superimposed jumps. 

The population size $N(t)$ is assumed to evolve according to the 
Gompertzian growth equation (1) between sudden decrements of magnitude $\epsilon N(t)$ with $\epsilon >0$, which occur at the event times of a Poisson process $N^*(\lambda; t)$ which has a constant rate
parameter $\lambda$.  The stochastic equation for $N(t)$ is thus 
\be  dN(t) = rN(t)(\ln K - \ln N(t))dt - \epsilon N(t)dN^*(\lambda; t), \ee
where $r$, $\epsilon$ and $K$ are all positive. The initial population size 
is $N_0\in (0,K]$. Note that if $\epsilon=1$ then the first event in the
Poisson process will annihilate the population so that it will be assumed that $0 <\epsilon <1$. This means that the population can never 
attain the value zero, but it may become arbitrarily close to zero. 

Let $p(N,t|N_0)$ be the transition probabiluty density function of $N$.
From Gihman and Skorohod (1972) we find that $p$ satisfies the forward Kolmogorov equation 
\be \frac{\p p}{\p t} =  \frac{\p }{\p N } \big[rN(\ln N - \ln K)p\big] +
 \lambda \big[p(N(1+\epsilon),t) - p(N,t)\big]. \ee
The moments of $N(t)$ can be found from this equation but closed form expressions do not seem to be available.

The principal objective is to determine how long it will take for the  population to attain a certain small value, suxh as unity, for a given 
initial size $N_0$ and given remaining parameters $r$, $\lambda$ and
$K$? In the case of a tumor or other aggregation of cells, we are 
asking for the time to eradicate the mass of cells or to greatly reduce its number.

Put $x=N_0$; now regarded as a variable. Then define the random variable $T(x)$ as the time to reduce the population to a level less than unity, for example.  That is,
\be T(x)= \inf\{t | N(t) \notin (1,K) |N_0=x \in (1,K]\},  \ee
which is the time that the number of cells first escapes from the interval $(0,K)$ which must occur at $N<1$ because $N$ cannot exceed $K$.
If the number of cells is $N(t)$ then $T(x)$ is the actual time of complete eradication because having less than 1 cell is equivalent to having no cells at all. Letting the $n-th$ moment of $T(x)$ be $M_n(x), n=0,1, 2,...$
so that
\be M_n(x) = E[T^n(x)], \ee
we find, from Tuckwell (1976), that these quantities satisfy the recurrence system of differential-difference equations,
\be rx(\ln K - \ln x) \frac{dM_n(x)}{dx} + \lambda [M_n(x(1-\epsilon)) -M_n(x) ] = -nM_{n-1}(x), \ee
for $n=1,2,...$ with $M_0(x)=1$, which is the probability that $N$ ever escapes from $(1,K)$.  For $n \ge 1$ the boundary conditions are that
$M_n(x)=0$ for $x \notin (1,K)$ and we also have the requirement that
$M_n(x)$ is bounded and continuous on $(1,K)$. Note, however that this system of differential-difference equations is singular at $x=K$.

Rather than solving Equ. (22) directly, it is convenient to note that the
simple transformations of Equ. (6) for the deterministic problem will reduce the stochastic problem to  a simpler one
for which the author and coworkers have obtained solutions, both 
analytical and numerical, in the context of the problem of determining the time interval between impulses of a model neuron.  The transformations result in the simple SDE 
\be dY(t)= -rY(t)dt + \alpha dN^*(\lambda; t), \ee
which describes a process which decays exponentially towards zero
between upward jumps of magnitude $\alpha$.  The time at which
the original process $N^*$ declines below unity for the first time is the same as the time at which $Y$ first exceeds the value $k$. This exit time is precisely the same as the time between impulses in a Stein (1965)
model neuron with a time constant $1/r$ and threshold $k$ which receives Poisson excitatory postsynaptic potentials with rate $\lambda$
and amplitude $\alpha$.  This threshold crossing problem for a discontinuous Markov process has been much studied (see Tuckwell 1975 and 1988, and references therein).

\subsection{Results}

Further simplifications are made by putting $Z=Y/\alpha$, so that
$Z$ has jumps of unit magnitude, and scaling time by
$\tau=\lambda t$ so the Poisson process has a rate parameter of unity.
Thus
\be dZ(\tau) = - \gamma Z(\tau) d\tau + d\hat{N}(\tau) \ee
where $\hat{N}$ has unit rate and $\gamma=r/\lambda$.   Defining
$\kappa= k/\alpha$ we find that the exit time of interest in (20) is
now 
\be \Theta (z) = inf\{ \tau | Z(\tau) > \kappa |Z(0)=z\}, \ee
whose moments $\mu_n(z)= E[\Theta^n(z)] $ satisfy the 
relatively simple system of equations 
\be -\gamma z  \frac{d\mu_n(z)}{dz} + \mu_n(z+1) - \mu_n(z) =
 - n\mu_{n-1}(z), \ee
for $n=1,2,...$. Here $z \in (0, \kappa)$ and for $n \ge 1$, $\mu_n(z)=0$
for $z$ outside this interval. 

Explicit expressions for the solutions of Equ. (26) with $n=1$ can 
be obtained for values of $\kappa$ between 0 and 3 when $\gamma$
takes on integer or fractional values (see for example Tuckwell and Richter, 1978). These results give the expectation of the time at which
$Z$ first reaches or exceeds $\kappa$ for an initial value $z$. For larger values of $\kappa$ numerical methods have been devised to solve the differential-difference equation.  In one such approach the differential-difference equation was converted to a system of ordinary differential equations on the unit interval (Tuckwell and Richter, 1978).  A different approach was employed in Cope and Tuckwell (1979) whereby an asymptotic expansion at large negative $z$ was matched to the
continued solution obtained with boundary conditions at $z=\kappa$
by means of a set of recursion relations.  Results for the process $Z$
can be readily converted to corresponding results for the original Gompertzian growth process $N$ because the transformations from
$Z$ to $N$ are monotonic.

There are two principal questions we will address. 
\begin{itemize}
\item{How does the extinction time depend on the size of the
population whose eventual size would be $K$ in the absence of decrements, for fixed values of the parameters $r$, $\lambda$ and 
$\epsilon$?}
\item{How does the extinction time depend on the various parameters
for different $K$ when the population is initially fully grown or almost
fully grown?}
\end{itemize}

Since there are so many combinations of parameters to explore, 
which makes it difficult to display results for them all, only a few representative cases will be reported here.   In the future, tables will be
published for the moments of the first exit time of $Z$ from various intervals. From these 
results for the extinction times of Gompertzian populations can be 
readily obtained. 

\begin{itemize}
\item {The first set of results, given in Figure 5, is designed to see how the
extinction time changes as a function of mean frequency and magnitude of the random decrements. The logarithm of the expectation of the extinction time $T$ for a population at saturation level is plotted against the
mean frequency of decrements (in units of $r$) for two values of the 
decrement $\epsilon=0.6321$ and $\epsilon=0.3935$ for $Z$, which two numbers
correspond to jump fractions of $\alpha=1$ and $\alpha=0.5$ in the original population $N$. The unit of the extinction time $T$ (corresponding to $T(K)$ in Equ. (20)) is $1/\lambda$. 

In the numerical example given here, $K$ is chosen to be $e^8 \approx 2981$ individuals or cells and $T$ is the time taken for the population  to become less than 1 individual or cell. From Figure 1 it can be seen that the waiting time for extinction ($<1$) is extremely large for small and moderate frequencies of decrements and that the logarithm of $E[T]$ goes about like $\exp(-\lambda/r)$ so that $E[T]$ goes about like $\exp(e^{-\lambda/r})$.
Note that the values of $\epsilon$ here correspond to relatively large decrements as of order half the population is removed at one hit. 

At a given $\lambda/r$ the effect of changing $\epsilon$ is exceedingly pronounced.  For example, with $\lambda=8r$, when $\epsilon$
goes from 0.63 to 0.39, $E[T]$ increases from $2.2/\lambda$
to $337.9/\lambda$!   Note that by scaling, the extinction time
for $K=e^8$ and $\epsilon=0.3935$ are the same as for $K=e^{16} \approx 8.9 \times 10^{16}$ and $\epsilon=0.6321$. 
}

     \begin{figure}[!h]
\begin{center}
\centerline\leavevmode\epsfig{file=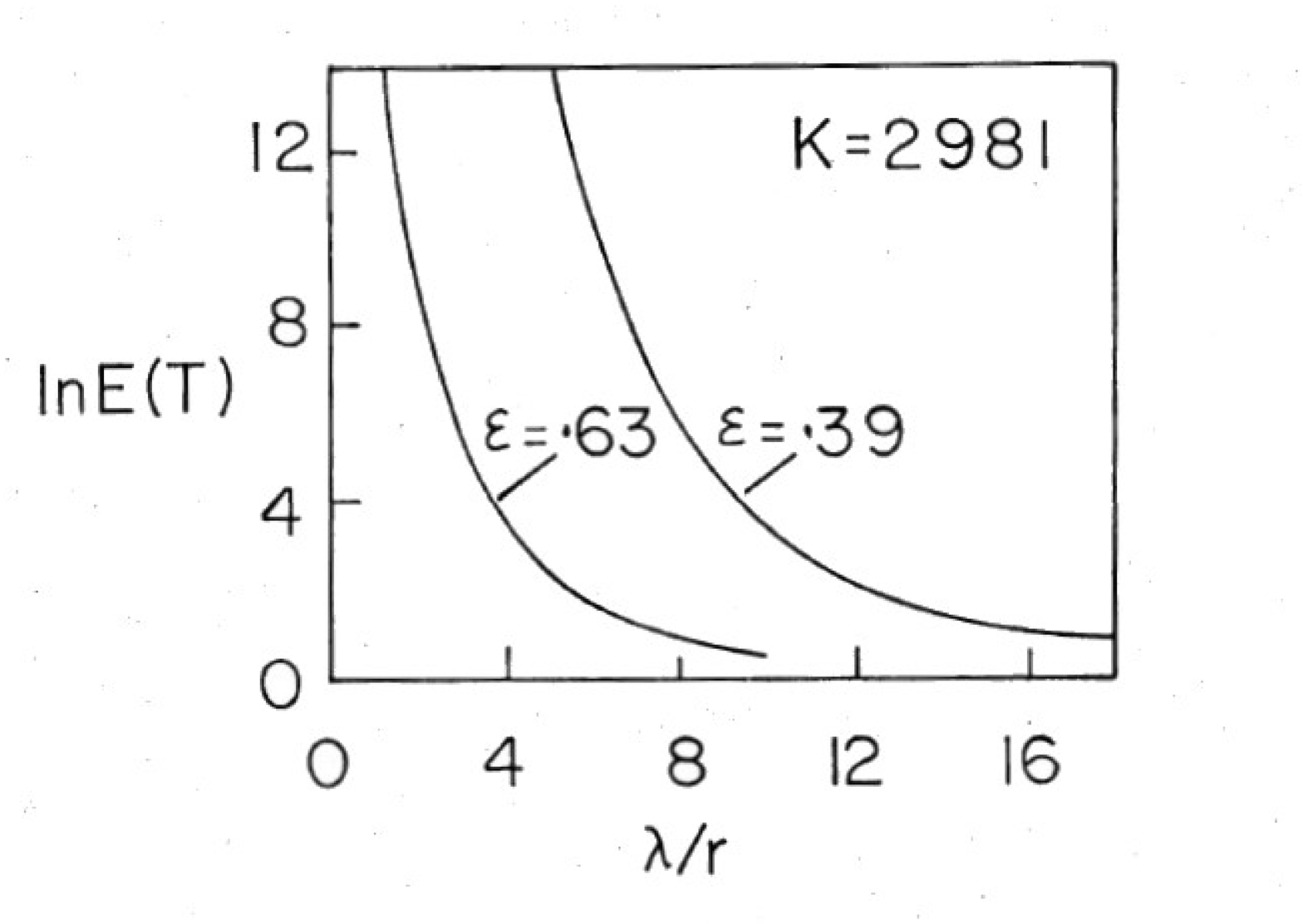,width=3.25in}
\end{center}
\caption{The dependence of the logarithm of the expected time
to extinction of a population of size $K=e^8$ with Poisson decrements
as a function of their mean rate. Results are shown for cases where
each hit removes fractions of $\epsilon=0.63$ and $\epsilon=0.39$
of the total population. The unit for the extinction time is $1/\lambda$ where $\lambda$ is the mean rate. } 
\label{fig:wedge}
\end{figure}

\item{Next we address the question of how much longer a larger 
population of individuals or cells will take to eradicate than a smaller one. To this end we plot in Figure 6, for fixed size of decrements $\epsilon=0.63$, and for various mean rates of their arrival $\lambda=r, 2r, 3r, 4r$ and $5r$, the 
logarithm of the expectation of the extinction time, in units of $1/\lambda$                against the logarithm of the total population size $K$. For each value of $\lambda$ the dependence of $\ln E[T]$ is approximately linear for small to moderate $K$, but for larger $K$ the growth is more of an exponential character.  In one set of results, for $\lambda=3r$,  when $K$ increases from 1000 to 2000 the expected extinction time increases from $49.4/\lambda$ to $109.9/\lambda$ which entails an approximate doubling in the extinction time for a doubling of the total population size.}

     \begin{figure}[!h]
\begin{center}
\centerline\leavevmode\epsfig{file=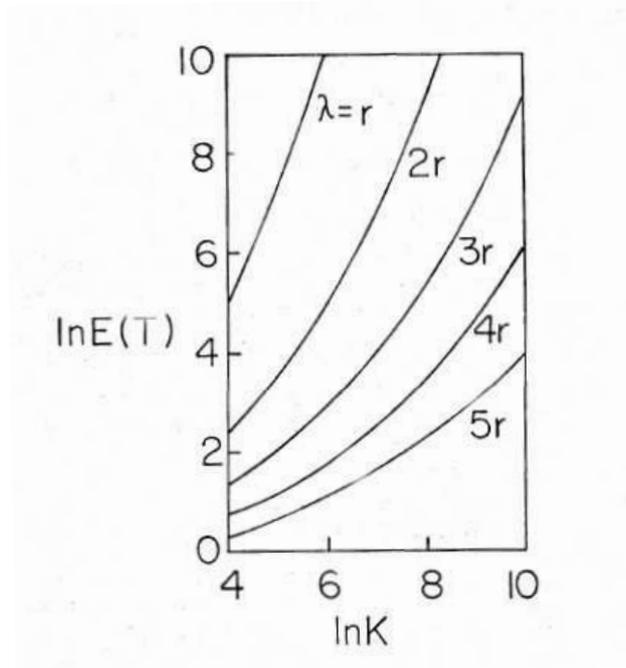,width=3.25in, angle=1}
\end{center}
\caption{The logarithm of the expectation of the extinction time
is plotted against maximum population size $K$ for various
mean frequencies of downward jumps. Each jump removes
a fraction 0.63 of the existing population. Units for $T$ as in Figure 1.
  } 
\label{fig:wedge}
\end{figure}

\item{Another question to address is how the extinction time
varies if the destructive decrements of the population commence at 
the early or late stages of growth. Hence we examine the variation in
the mean extinction time as the initial population size changes. 
An illustrative example is depicted in Figure 7. A population of organisms or cells whose eventual size would be $K=e^6$ is subjected to random hits
which remove a fraction 0.63 of the population. The mean time to
extinction is plotted as a function of the logarithm of the initial 
population size $N_0$ for two mean rates of hits of $\lambda=5r$ and
$\lambda=10r$. The curves are drawn through points 
 at which the logarithm of $N_0$ is an integer. 

The dependence of $E[T]$ on $\ln N_0$ is quite gentle for the
higher frequency of decrements but when the frequency is
$\lambda=5r$ we see that the extinction time increases rather sharply
for small $\ln N_0$. This indicates that if the removals of masses of the population starts to occur when the population is quite small, then long extinction times will be avoided and the population will quickly vanish. }

     \begin{figure}[!h]
\begin{center}
\centerline\leavevmode\epsfig{file=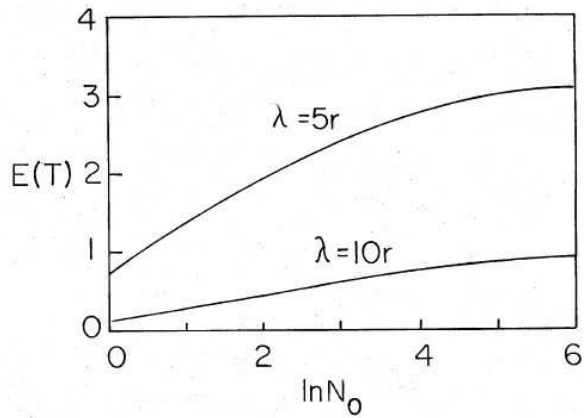,width=3.25in}
\end{center}
\caption{The expected extinction or eradication time as a function of
the population size when the downward jump process starts, for two values of the jump rate. The population has a maximum possible size
of $e^6$ individuals.} 
\label{fig:wedge}
\end{figure}

\item{Finally, the variability of the extinction time can be found
by calculating the second moment of $T$ from Equ. (22) or equivalently Equ. (26), provided the first moment has already been calculated. In Figure 8 is shown a plot of the logarithm of the standard deviation of the time to extinction of a population at saturation level $K=e^8 \approx 2981$ individuals or cells as a function of the mean arrival rate of decrements for two values of the fraction $\epsilon$ removed by each hit.  These curves give the standard deviations for the mean extinction times plotted in Figure 5. The dependences of the mean and standard deviation on mean arrival of hits are very similar.  For $\epsilon=0.63$
the coefficient of variation (CV, standard deviation/mean) is in fact very close to unity, indicative of a wide spread in the distribution,  until $\lambda/r \approx 3$ and decreases steadily to about 0.49
by $\lambda/r \approx 10$.  For $\epsilon=0.39$ the CV is near unity until $\lambda/r \approx 8$ whereupon it
decreases to attain the value 0.42 by $\lambda/r \approx 18$.}
\end{itemize}

     \begin{figure}[!h]
\begin{center}
\centerline\leavevmode\epsfig{file=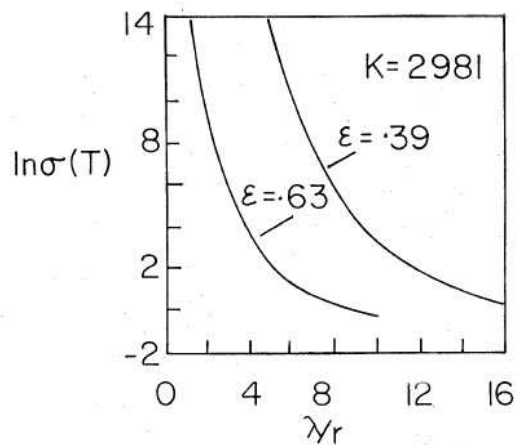,width=3.25in, angle=0.5}
\end{center}
\caption{The logarithm of the standard deviation of the extinction time
for the cases for the means in Figure 5.} 
\label{fig:wedge}
\end{figure}

%
%
%
%
%

\section{Summary and conclusions}
Gompertz models have been found to fit
well the  growth of many tumors and have also been
considered for certain animal and other populations of organisms.
Many articles have addressed the responses of tumors to radiation therapy and chemotherapy, each kind of therapy requiring a different
mathematical treatment. In this article the results of removing a fraction 
of the population at periodic intervals is explored by means of exact
results obtained by solving the assumed underlying differential equation.
Having defined a standard parameter set, the effect of varying, one at a time, the 4 key parameters of fraction removed ($\epsilon$ or $\alpha$), period ($T$), initial population size ($N_0$) and intrinsic growth rate ($r$) were examined by examples presented in Figures 1 to 4. Surprisingly, as seen in Figure 3, 
after about 15 decrements (over 15 periods) the resulting population size was practically
the same for a wide range of initial values. 
In the second part of this article the periodic occurrence of decrements
was replaced by their occurring at event times in a Poisson process.
The resulting discontinuous stochastic process was easily transformed to a simpler one which had been used as a model for nerve firings.
Hence results from the interspike interval calculations could be applied to the case of Gompertziian growth with Poissonian downward jumps.
The transition density of the process was not found but first and second order moments could be obtained from previous studies.
The coefficient of variation of the extinction time was near unity for 
large ranges of smaller values of mean input frequency.

\section*{Acknowledgements}
I am grateful to Professor Charles E. Smith of NCSU for introducing me
to the tumor growth literature,

\section*{References}

\nh Ai, B.Q., Wang, X.J., Liu, G.T. and Liu, L.G., 2003. Correlated noise in a logistic growth model. Physical Review E, 022903.

\nh Aroesty, J., Lincoln, T., Shapiro, N, Boccia, G., 1973.
Tumor growth and chemotherapy: mathematical methods, computer simulations and experimental foundations. Math. Biosci. 17, 243-300. 

\nh Badri, H., Salari, E., Watanabe, Y. and Leder, K., 2016. Optimizing chemoradiotherapy to target multi-site metastatic disease and tumor growth. arXiv preprint arXiv:1603.00349.

\nh Bajzer, $\check{\rm Z}$, Vuk-Pavlovi\'c, S. and Huzak, M., 1997. Mathematical modeling of tumor growth kinetics. In A Survey of Models for Tumor-Immune System Dynamics (pp. 89-133). Birkhäuser Boston.
%

\nh Beever, E.A.,  Aldridge, C.L., 2011. Influences of free-roaming equids on sagebrush ecosystems, with focus on greater sage-grouse. Studies in Avian Biology 38, 273-290.

\nh Behera, A., O'Rourke, S.F.C., 2008. The effect of correlated noise in a Gompertz tumor growth model. Brazilian Journal of Physics 38, 272-278.

\nh Berger, R.D., 1981. Comparison of the Gompertz and Logistic Equations to Describe Plant Disease Progress. Phytopathology 71, 716-719.

\nh Bortfeld, T., Ramakrishnan, J., Tsitsiklis, J.N. and Unkelbach, J., 2015. Optimization of radiation therapy fractionation schedules in the presence of tumor repopulation. INFORMS Journal on Computing, 27(4), pp.788-803.

%
%

\nh Burton, A.C., 1966. Rate of growth of solid tumors as
 a problem of diffusion. Growth 30, 157-176.

\nh Cope, D.K., Tuckwell, H.C.,1979. Firing rates of neurons with random excitation and inhibition. J. Theor. Biol. 80 , 1-14.

\nh Davies, G.R., 1927. The growth curve. J. Amer. Stat. Assoc. 22, 370-374.

\nh Dibrov, B.F., Zhabotinsky, A.M., Neyfakh, Y.A., Orlova, M.P. and Churikova, L.I., 1985. Mathematical model of cancer chemotherapy. periodic schedules of phase-specific cytotoxic-agent administration increasing the selectivty of therapy. Mathematical biosciences, 73(1), pp.1-31.
%
%

\nh Eberhardt, L.L., Breiwick, J.M., 2012. Models for population growth curves. ISRN Ecology, 2012.

\nh Feller, W., 1940.  On the integro-differential equations of purely
discontinuous Markoff processes. Trans. Amer. Math. Soc. 48, 488-515.


%
%
%

%
%

\nh Ferguson, J.M., Ponciano, J.M., 2015. 
Evidence and implications of higher-order scaling
in the environmental variation of animal
population growth. PNAS 112, 2782-2787. 

\nh Gihman, I.I., Skorohod, A.V., 1972. Stochastic differential equations. Springer, Berlin. 

\nh Gompertz, B., 1825. On the nature of the function expressive of the law of human mortality, and on a new mode of determining the value of life contingencies. Philosophical transactions of the Royal Society of London 115, 513-583.

\nh Ito, K., 1951. On stochastic differential equations. Mem. Amer. Math. Soc. 4.  

\nh Jansson, B., Revesz, L., 1974. Analysis of the growth of tumor cell
populatiopns. Math. Biosci. 19, 131-154.

%

%

\nh Laird, A.K., 1964. Dynamics of tumour growth. British J. Cancer 18, 490-502.

\nh Lefever, R., Horsthemke,W., 1979.  Bistability in fluctuating environments.
Implications in tumor immunology. Bull. Math. Biol. 41, 469-490.

%

%
%

\nh Murphy, H., Jaafari, H., Dobrovolny, H.M., 2016.
Differences in predictions of ODE models
of tumor growth: a cautionary example. BMC Cancer 16, 163-172.

\nh Norton, L., Simon, R., Brereton, H.D., Boyden, A.E., 1976.
Predicting the course of Gompertzian growth. Nature 264, 542-545.

\nh Norton, L., 1988. A Gompertzian model of human breast cancer growth. Cancer research, 48(24 Part 1), pp.7067-7071.
%

\nh Oksendal, B., 2000. Stochastic Differential Equations, 5th Edn. 
Springer, Berlin.

\nh Rockne, R., Alvord Jr, E.C., Rockhill, J.K., Swanson, K.R., 2009. A mathematical model for brain tumor response to radiation therapy. J. Math. Biol. 58, 561-578.

\nh Sachs, R.K., Hlatky, L.R., Hahnfeldt, P., 2001. Simple ODE models of tumor growth and anti-angiogenic or radiation treatment. Mathematical and Computer Modelling 33, 1297-1305.

\nh Salari, E., Unkelbach, J. and Bortfeld, T., 2015. A mathematical programming approach to the fractionation problem in chemoradiotherapy. IIE Transactions on Healthcare Systems Engineering, 5(2), pp.55-73.

%

\nh Simpson-Herren, L., Lloyd, H.H., 1970. Kinetic parameters and
growth curves for experimental tumor systems. Cancer Chemotherapy
Reports Part 1 54, 143-174. 

\nh Skorohod, A.V., 1965. Studies in the theory of random processes.
Addison-Wesley, Reading, Mass.

\nh Smith, C.E., Tuckwell, H.C., 1974.  Some stochastic growth processes. In: Mathematical Problems in Biology,
Springer, Berlin,  pp 211-225.

 \nh Tuckwell, H.C., 1974. A study of some diffusion models of
population growth. Theor. Pop. Biol. 5, 345-357.

\nh Tuckwell, H.C., 1975. Determination of the inter-spike times of neurons receiving randomly
arriving post-synaptik potentials. Biol. Cybernetics 18, 225-237. 

\nh Tuckwell, H.C., 1976. On the first-exit time problem for temporally
homogeneous Markov processes. J. Appl. Prob. 13, 39-48. 

\nh Tuckwell, H.C., 1988. Introduction to theoretical neurobiology, Volume 2. Cambridge University Press, Cambridge, UK. 

\nh Tuckwell, H.C., Le Corfec, E., 1998. A stochastic model for early HIV-1 population dynamics. J. Theor. Biol. 195, 451-463. 

\nh Tuckwell, H.C., Richter, W., 1978.  Neuronal interspike time distributions and the estimation of neurophysiological and neuroanatomical parameters. J. Theor. Biol. 71, 167-183.

%

%
%
%

\nh Vaidya, V.G.,  Alexandro, F.J., 1982. Evaluation of some mathematical models for tumor growth. International journal of biomedical computing, 13, 19-35.

\nh Winsor, C.P., 1932. The Gompertz curve as a growth curve. PNAS 18, 1-8.

\nh Zullinger, E.M., Ricklefs, R.E., Redford, K.H., Mace, G.M., 1984. Fitting sigmoidal equations to mammalian growth curves. J. Mammalogy 65, 607-636.

\end{document}